\documentclass[letter]{aa}
\usepackage{epsfig}
\usepackage{amsmath}
\usepackage{subfigure}
\usepackage{natbib}
\usepackage{ulem}
\usepackage{multirow}
\usepackage{color}
\usepackage{longtable}
\usepackage{graphicx}
\usepackage{epstopdf}
\usepackage{booktabs}
\usepackage{txfonts}

\newcommand{\jmst}{J.~Mol.~Struct.}   
\newcommand{\jms}{J.~Mol.~Spectrosc.}

\begin{document}   

\title{Detection of the linear SiC$_3$ and SiC$_5$ radicals in 
IRC\,+10216\thanks{Based on observations carried out with the 40m radio telescope 
of the Yebes Observatory (projects 19A010, 20A017, 20B014, 21A019, and commissioning observations), operated by the Spanish Geographic Institute  (IGN, Ministerio de Transportes, Movilidad y Agenda Urbana).
IRAM is supported by INSU/CNRS (France), MPG (Germany) and IGN (Spain).}}

\author{
J.~Cernicharo\inst{1}
J.~R.~Pardo\inst{1},
M.~Ag\'undez\inst{1},
J.P. Fonfr\'{\i}a\inst{1},
L. Velilla-Prieto\inst{1},
C.~Cabezas\inst{1},
B.~Tercero\inst{2,3},
P.~de Vicente\inst{3},
\and 
M.~Gu\'elin\inst{4},
}

\institute{Consejo Superior de Investigaciones Científicas, Instituto de F\'isica Fundamental, C/ Serrano 121, 28006 Madrid, Spain
\newline \email jose.cernicharo@csic.es \& jr.pardo@csic.es
\and Observatorio Astron\'omico Nacional (OAN, IGN), Calle Alfonso XII 3, 28014 Madrid, Spain.
\and Centro de Desarrollos Tecnol\'ogicos, Observatorio de Yebes (IGN), 19141 Yebes, Guadalajara, Spain.
\and Institut de Radioastronomie Millim\'etrique, 300 rue de la Piscine, F-38406, Saint Martin d'H\`eres, France
}

\date{Received 26 July 2025; accepted 6 August 2025}

\abstract{
We detected the linear $^3\Sigma^-$ radicals SiC$_3$ and SiC$_5$ toward IRC+10216
using an ultrasensitive line survey gathered with the Yebes 40\,m radio telescope.
The derived column densities of $l$-SiC$_3$ and $l$-SiC$_5$
are (3.6$\pm$0.4)$\times$10$^{12}$ cm$^{-2}$ and (1.8$\pm$0.2)$\times$10$^{12}$ 
cm$^{-2}$, 
respectively. The linear SiC$_3$ radical is $\sim$2 times less abundant that its 
singlet rhomboidal prolate
isomer, for which we provide a new analysis based on recent sensitive observations
in the
Q band (7\,mm), and at 3 and 2\,mm with the IRAM 30m telescope. 
The emission detected from these species arises from the cool external layers 
of the circumstellar
envelope. We speculate whether ion-neutral routes involving SiC$_n$H$_m$$^+$ 
cations or neutral-neutral 
reactions involving Si and SiC$_2$ could efficiently synthesize these species.}


\titlerunning{Detection of $l$-SiC$_3$ and SiC$_5$ in IRC+10216}
\authorrunning{Cernicharo et al.}

\maketitle

\section{Introduction}
The search for chemical complexity in space has received a significant boost in recent years
thanks to the ultrasensitive line surveys in the Q band of the starless core TMC-1, the QUIJOTE\footnote{\textbf{Q}-band \textbf{U}ltrasensitive \textbf{I}nspection \textbf{J}ourney to the \textbf{O}bscure \textbf{T}MC-1 \textbf{E}nvironment} line survey
\citep[see, e.g.,][]{Cernicharo2021a,Cernicharo2021b,Cernicharo2023a,Cernicharo2024},
and the envelope (IRC+10216) of the carbon-rich star CW Leo 
\citep[see, e.g.,][]{Pardo2021,Pardo2022,Pardo2025a,Changala2022,Cernicharo2023b,
Cabezas2023,Gupta2024}. Since 2020 more than 90 molecules have been found with the Yebes 40m radio telescope
thanks  to these ultrasensitive line surveys.

Since the detection of SiC$_2$ in 1984 by Thaddeus et al., this species has been observed toward the envelopes of carbon-rich stars through
its rotational transitions in the millimeter and submillimeter domains. 
These lines provide information on the whole
circumstellar envelope of these objects \citep{Lucas1995,Cernicharo2010,Cernicharo2018,Velilla2023}.
The spatial distribution of SiC$_2$ derived from these high excitation lines is only known in detail toward IRC\,+10216, where it was observed with the Plateau de Bure interferometer
\citep{Guelin1993,Velilla2019} and the CARMA 
and ALMA interferometers \citep{Fonfria2014,Velilla2015,Velilla2023}. 
Observations of three C-rich stars in four low-energy SiC$_2$ transitions at an angular resolution of  $\sim$1$''$ revealed shell-like structures 
\citep{Feng2023}.
A study with the IRAM 30\,m telescope showed that SiC$_2$ is less abundant in denser envelopes, suggesting efficient incorporation of SiC$_2$ into dust grains \citep{Massalkhi2018}.

Apart from SiO, SiS, SiC$_2$, and Si$_2$C \citep{Cernicharo2015}, only a few 
molecules containing silicon have been found, all of which are present in IRC+10216. 
Silicon carbide (SiC) was detected in
the external shells (300 R$_*$ and beyond) of IRC\,+10216 
\citep{Cernicharo1989,Patel2013,Velilla2019}.  
It was also detected in the envelopes of several 
C-rich asymptotic giant branch stars by \citet{Massalkhi2018}.
Other silicon carbides such as the rhomboidal SiC$_3$
\citep{Apponi1999a}, SiC$_4$ \citep{Ohishi1989}, and SiC$_6$ \citep{Pardo2025a} have been found in this source with
column densities around 10-100 times lower than those of SiC$_2$ and Si$_2$C. SiH$_4$ was detected in the infrared by \citet{Keady1993} 
and found to be formed at distances larger than 80 stellar radii
\citep{Monnier2000}.
Additional molecules containing
silicon in this source are SiN \citep{Turner1992}, SiH$_3$CH$_3$, and SiH$_3$CN \citep{Cernicharo2017}, and the two isomers SiNC and SiCN
\citep{Guelin2000,Guelin2004}, all in low abundances.

The detection of long SiC$_n$ chains is expected based on the abundances found for the
smaller members. 
In this Letter we report the detection of the linear $^3\Sigma^-$ radicals SiC$_3$ and SiC$_5$
(hereafter referred to as $l$-SiC$_3$ and $l$-SiC$_5$). Linear SiC$_3$ is an
isomer of the already detected rhomboidal closed shell species SiC$_3$ \citep[hereafter referred to as $r$-SiC$_3$;][]{Apponi1999a}. 
The abundance of the newly detected molecules are compared to those of the other members
of the SiC$_n$ family and discussed in the context of the possible reactions leading to their
formation.

\begin{figure}[t] \begin{center}
\includegraphics[width=0.495\textwidth]{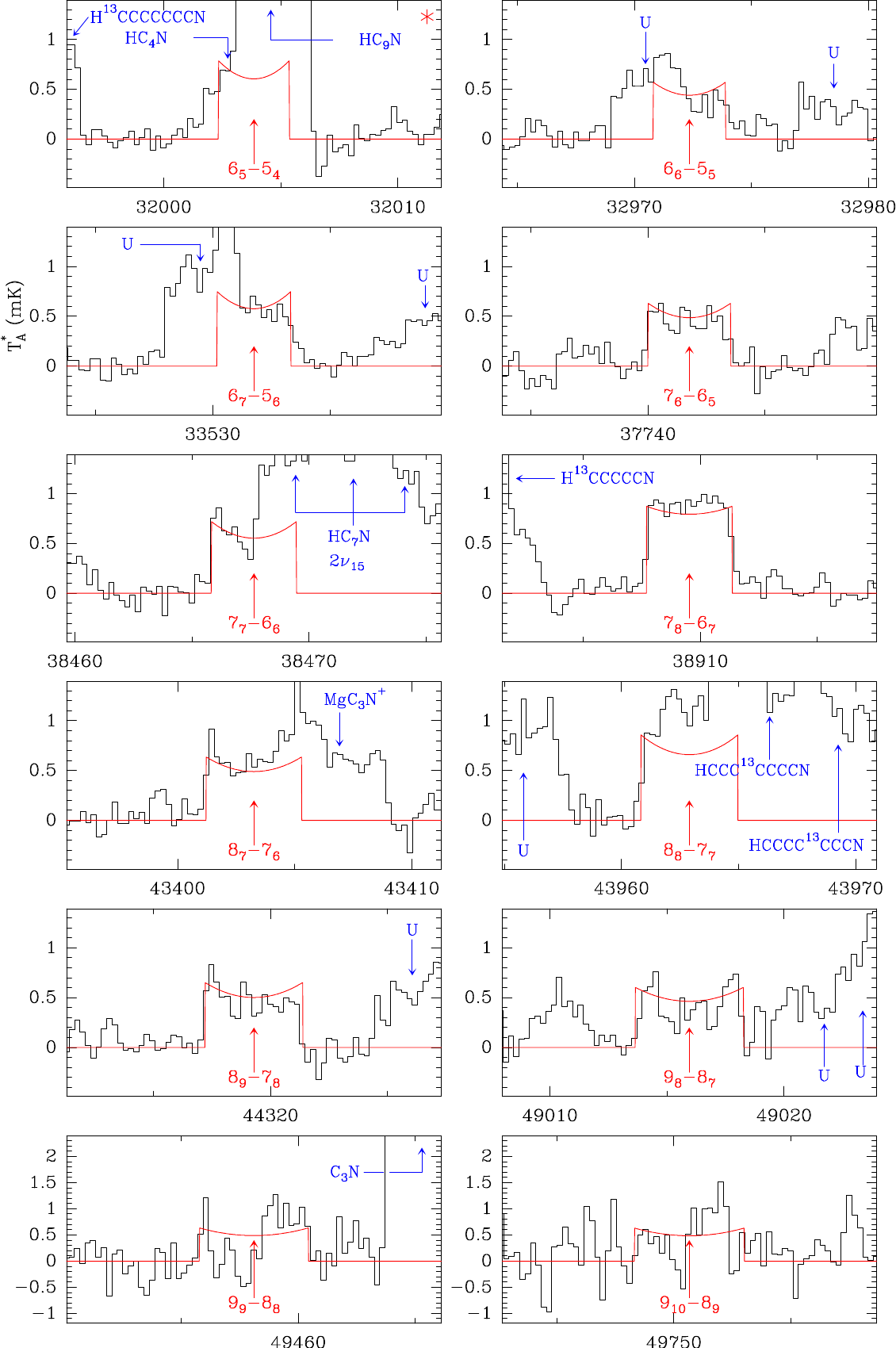}\end{center}  
  \caption{Lines of the linear $^3\Sigma^-$ $l$-SiC$_3$ radical observed in this work, with their fitted line
  profile shown in red. Features from other species are labeled in blue. The abscissa corresponds
    to the rest frequency (in MHz) adopting a v$_{LSR}$ of $-$26.5 km\,s$^{-1}$ \citep{Cernicharo2000}.
    The ordinate corresponds to the antenna temperature corrected for telescope losses and atmospheric attenuation
    (in mK). Fully blended lines are indicated by a red star in the upper-right corner 
        of the first panel. 
        Quantum numbers are $N$ (rotational quantum
number) and $J$ (total angular momentum, $J$ = $N$, $N$$\pm$1). They are
written as $N_J$.
        }         
    \label{fig_sic3}
\end{figure}

\section{Observations} 
\label{sect:obs}
The data presented in this Letter come from a still ongoing
31-50 GHz spectral survey of IRC+10216 
($\alpha_{J2000}$=09$^h$ 47$^m$ 57.36$^s$, $\delta_{J2000}$=+13$^o$ 16' 44.4$''$)
gathered in the framework of the Nanocosmos                                     
project\footnote{\texttt{https://nanocosmos.iff.csic.es/}}.
The data were obtained 
with the 40 meter antenna of Yebes Observatory (IGN, Spain). 
A total
of 1360 hours of on-source telescope time was acquired between
April 2019 and September 2024. The experimental setup is described in                                       
detail in \citet{Tercero2021}.  

The observing mode was position-switching with an off position at 300$''$ in azimuth.                  
Pointing corrections were obtained by observing the SiO masers 
of the nearby star R Leo, and errors were always            
 within 5$''$. The intensity scale of the final calibrated data is antenna temperature (T$_A^*$) corrected for atmospheric absorption using the ATM package                            
\citep{Pardo2001,Pardo2025b}. Absolute calibration  
uncertainties are estimated to be within 10~\%.
Although the primary spectral resolution is 38.1 kHz, the final spectrum
was obtained by applying a six-channel box smoothing to the nominal 
spectrum. Hence, the spectral resolution is $\Delta\nu$=229 kHz. 
This line survey is 50-100 times more sensitive than previous surveys of this source at
the same frequencies obtained with the Nobeyama 45 m radio telescope \citep{Kawaguchi1995}. 
The observations with the 30m IRAM radio 
telescope have been described in \citet{Cernicharo2019}.

\begin{figure*}[t] \begin{center}
\includegraphics[width=0.98\textwidth]{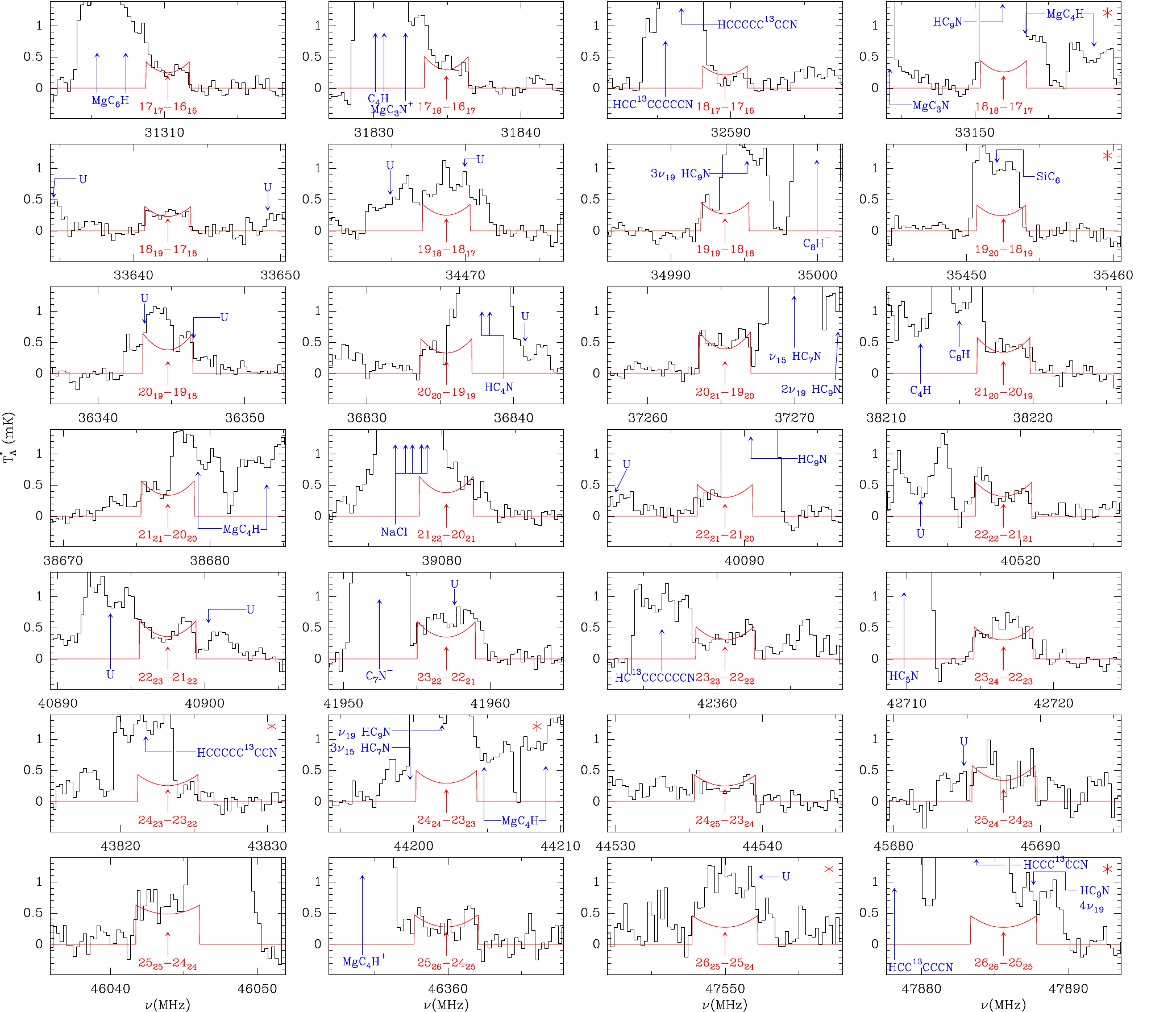}\end{center}  
  \caption{Lines of $l$-SiC$_5$ observed in this work.
   Features from other species are labeled in blue. Fully blended lines are
    indicated by a red star in the upper-right corner of the corresponding panel. 
        Fitted line profiles, abscissa, ordinate, and quantum numbers
    are as in Fig. \ref{fig_sic3}.
        }         
   \label{fig_sic5}
\end{figure*}

\section{Results}
\label{sec:results}
Line identification was achieved using the MADEX, CDMS,
and JPL catalogs \citep{Cernicharo2012,Muller2005,Pickett1998}.
The datasets were fitted using the SHELL method of the 
GILDAS\footnote{\texttt{http://www.iram.fr/IRAMFR/GILDAS}} package. The global fitting procedure 
described by \citet{Pardo2025a} was adopted to derive the integrated intensities of the
lines. 
Their typical U-shaped profiles, together with the expanding velocity  of the envelope (14.5 km\,s$^{-1}$; \citealt{Cernicharo2000}), enabled the detection of very weak lines emerging from IRC+10216.

Rotational line frequencies for $l$-SiC$_3$ and $l$-SiC$_5$ are available from 
laboratory experiments by \citet{McCarthy2000}. Both species have a $^3\Sigma^-$ electronic ground state.
A dipole moment of 4.8\,D was calculated for $l$-SiC$_3$ in the same
work. For $l$-SiC$_5$ we have assumed a dipole moment of 6.6\,D \citep{Muller2005}.
Both species have been implemented in MADEX \citep{Cernicharo2012}. The frequency coverage and
the large quantum numbers for the laboratory transitions guarantee very 
good frequency predictions, with uncertainties below a few kilohertz, in the Q band. The observed lines of 
$l$-SiC$_3$ are shown in Fig. \ref{fig_sic3}, and those of $l$-SiC$_5$ are
shown in Fig. \ref{fig_sic5}. In spite of the presence of some lines blended 
with other features, the
detection of $l$-SiC$_3$ is based on the detection of 11 lines,
and that of $l$-SiC$_5$ in 22 lines. For both species, only a few lines are 
fully blended with strong features. 
These transitions were not considered for the analysis. Hence, the detection appears robust and permits a detailed analysis of the abundance of these species.

We used the observed integrated intensities given in Table \ref{line_parameters} to create rotation 
diagrams. For $l$-SiC$_3$ we derive {T$_{rot}$=5.2$\pm$0.5\,K and N=(3.6$\pm$0.5)$\times$10$^{12}$ cm$^{-2}$.
For $l$-SiC$_5$ the results are T$_{rot}$=17.2$\pm$1.6\,K and N=(1.8$\pm$0.3)$\times$10$^{12}$ cm$^{-2}$.
These column densities are between those derived by \citet{Pardo2025a}
for SiC$_4$ (N=(7.3$\pm$1.1)$\times$10$^{12}$ cm$^{-2}$; T$_{rot}$=21.5$\pm$5\,K)
and SiC$_6$ (N=(7.9$\pm$1.5)$\times$10$^{11}$ cm$^{-2}$; T$_{rot}$=19.5$\pm$2.1\,K). 

Laboratory spectroscopy
for rhomboidal SiC$_3$ was obtained by \citet{McCarthy1999a} and \citet{Apponi1999b}. Its column density in IRC+10216
was derived by \citet{Apponi1999a} from several lines in the millimeter domain to be
4\,$\times$\,10$^{12}$ cm$^{-2}$. No new data on $r$-SiC$_3$
have been published since then. Hence, we analyze in Appendix 
\ref{app_r-SiC3} the data we
have obtained in the 7\,mm (Q band), 3\,mm, and 2\,mm frequency 
domains for this species. Two temperature components are found
(see Fig. \ref{fig_trot_r-sic3}), one with
T$_{rot}$=15.8$\pm$0.9\,K and N=(6.5$\pm$0.6)$\times$10$^{12}$ 
cm$^{-2}$. The rotational temperature
for this component is
similar to that found for $l$-SiC$_5$, SiC$_4$, and SiC$_6$. A second component is needed with a higher rotational temperature
of 44$\pm$8\,K and a lower column density of (2.1$\pm$0.5)$\times$10$^{12}$ cm$^{-2}$.
The fact that $r$-SiC$_3$ has a C$_{2v}$ symmetry with only $a$-type transitions (only $\Delta$$K_a$=0 transitions are allowed) introduces significant different excitation conditions from those of $l$-SiC$_3$ and $l$-SiC$_5$ (see Appendix \ref{app_r-SiC3}). These results, together with 
the U-shaped line profiles (see Fig. \ref{fig_r-sic3}), indicate that
these species are found in the largest abundances in the external layers. For $r$-SiC$_3$ the warm component with T$_{rot}$=44\,K
could indicate that the molecule starts to be produced in the intermediate layers of the envelope, or that infrared pumping is
playing a significant role in the high excitation lines of this molecule. 
Another isomer, the oblate cyclic
SiC$_3$ species, which is 4.7 kcal/mol above the prolate one \citep{Rintelman2006}, has also been observed 
in the laboratory 
\citep{McCarthy1999b}. However, only a few
lines were measured and, consequently, the frequency predictions are uncertain. Moreover, the dipole moment
of this isomer is a factor of 2 below that of the rhomboidal one. 
Hence, its lines are expected to be at least a factor
of 4 weaker than those of $r$-SiC$_3$ and $l$-SiC$_3$. Using the $3_{03}-2_{02}$ line, 
which has been measured in the laboratory  at 41641.985$\pm$0.005 MHz, 
and adopting the rotational temperature derived for $r$-SiC$_3$,
we
derive a 3$\sigma$ upper limit to the column  density of this isomer of 
7$\times$10$^{11}$ cm$^{-2}$.

\section{Discussion}\label{sec:discussion}

It is interesting to compare the column density of  rhomboidal SiC$_3$ (see Appendix \ref{app_r-SiC3}) 
with that of $l$-SiC$_3$. Despite the fact that the linear $^3\Sigma^-$ isomer
is 2.2 kcal/mol 
above the rhomboidal one \citep{Rintelman2006}, the low
rotational temperature components of the two isomers have similar column densities, $r$-SiC$_3$/$l$-SiC$_3\sim$1.8.
From the observed U-shaped line profiles of the $r$-SiC$_3$ transitions (see Fig. \ref{fig_r-sic3}),
it is clear that most of the emission is located in the external layers
of the envelope, where the kinetic temperature is $\sim$25\,K \citep{Guelin2018}. 
The low rotational temperature derived for $l$-SiC$_3$, T$_{rot}$$\sim$5.2\,K, also suggests a formation
in the cold external layers. The difference between the rotational temperatures of 
the linear ($^3\Sigma^-$) and rhomboidal 
($^1A$, C$_{2v}$ symmetry) isomers has to be related to the different structure 
and excitation conditions of the two species. 
Moreover, this low T$_{rot}$ is compatible with that derived for
species of similar size such as MgC$_2$ (6$\pm$1\,K; \citealt{Changala2022}) and
CaC$_2$ (5.8$\pm$0.6\,K; \citealt{Gupta2024}).

The chemistry of SiC$_3$ in IRC\,+10216 has been investigated to some extent. \cite{MacKay1999} predicted a sizable column density for SiC$_3$ (without distinction of the isomer) of 4.6\,$\times$\,10$^{12}$ cm$^{-2}$. Although not specifically discussed by these authors, in their model SiC$_3$ most likely arises from the photodissociation of SiC$_4$, which was predicted to be five times more abundant than SiC$_3$. Photodissociation of SiC$_4$ is, however, unlikely to be the main source of SiC$_3$ because the observed column density of SiC$_4$ (7.3\,$\times$\,10$^{12}$ cm$^{-2}$; \citealt{Pardo2025a}) is similar to the sum of those derived for $r$-SiC$_3$ (4.3\,$\times$\,10$^{12}$ cm$^{-2}$; \citealt{Apponi1999a}) and $l$-SiC$_3$ (3.6\,$\times$\,10$^{12}$ cm$^{-2}$; this study).

Other routes leading to SiC$_3$ in current astrochemical databases \citep{Millar2024,Wakelam2024} involve the dissociative recombination with electrons of Si-bearing hydrocarbon cations SiC$_n$H$_m^+$ formed by ion-neutral reactions involving Si or Si$^+$. The reaction between Si$^+$ and C$_2$H$_2$ has been studied experimentally, and it has been found that the cation SiC$_2$H$^+$ is formed \citep{Wlodek1991}. It would be interesting to investigate how Si$^+$ reacts with polyynes and cyanopolyynes and their radicals, 
particularly whether radiative association proceeds rapidly for long carbon chains, as observed for metals such as Mg$^+$, Na$^+$, and 
Al$^+$ \citep{Petrie1996,Dunbar2002,Cabezas2023,Cernicharo2023b}. The formation of cationic complexes of the type SiC$_n$H$_2^+$, SiC$_n$H$^+$, SiHC$_n$N$^+$, and SiC$_n$N$^+$ followed by their dissociative recombination with electrons could efficiently form the SiC$_n$ chains.

Neutral-neutral reactions could also allow SiC$_n$ chains to be synthesized. For example, SiC$_3$ could be formed in the reactions Si + C$_3$H and SiC$_2$ + CH, while SiC$_5$ could be produced by the reactions Si + C$_5$H, SiC$_2$ + C$_3$H, and SiC$_4$ + CH. If these reactions are barrierless and proceed through H atom elimination, they can efficiently form SiC$_3$ and SiC$_5$ because they involve reactants that are known or expected to be abundant in the outer circumstellar layers of IRC\,+10216. However, none of them have been studied either experimentally or theoretically. Reactions of Si with unsaturated closed electronic shell hydrocarbons have been measured to be rapid at low temperatures \citep{Canosa2001}, making it likely that Si reacts efficiently with C$_3$H and C$_5$H. 
We note that the reaction Si + C$_2$H$_2$ is endothermic for the 
channel SiCCH + H \citep{Kaiser2009}, which explains why SiCCH (whose rotational spectrum is known; \citealt{Apponi2000,McCarthy2001}) has not been detected in IRC+10216.
Other Si-bearing molecules with well-characterized rotational spectra that could result from the various ion-neutral or neutral-neutral routes at work in IRC\,+10216 are H$_2$CSi, H$_2$CCSi, H$_2$CCCSi, and H$_2$CCCCSi \citep{Izuha1996,McCarthy2002,McCarthy2024}. However, none of them is found in our data at 7, 3, and 2\,mm.

In summary, $l$-SiC$_3$ and $l$-SiC$_5$ have been found in IRC+10216, completing the family of
silicon carbon chains from SiC up to SiC$_6$.
The column densities of these species
are (3.6$\pm$0.4)$\times$10$^{12}$ cm$^{-2}$ and
(1.8$\pm$0.2)$\times$10$^{12}$ cm$^{-2}$, respectively. 
They are expected to be formed in the outer envelope.
The abundance ratio $l$-SiC$_3$ over $l$-SiC$_5$
is $\sim$2. 
These species are 17 and 33 times less abundant than SiC
\citep{Cernicharo1989} and have a similar 
abundance to that of SiC$_6$ \citep{Pardo2025a}.
We expect to detect heavier members of the SiC$_n$ family with future improved 
data of IRC+10216.

\begin{acknowledgements}
We thank the ERC for funding through grant ERC-2013-SyG-610256-NANOCOSMOS
This publication has been also funded 
by MICIU/AEI/10.13039/501100011033, FEDER and by ESF+
through grants PID2023-147545NB-I00, PID2022-137980NB-100, and RYC2023-045648-I. 
\end{acknowledgements}

\normalsize

\onecolumn
\begin{appendix}
\section{Line parameters}
\label{ap_line_parameters}
Although the data presented in this work in the Q band correspond to windows of $\pm$8 MHz around the
centre frequencies, the data were analysed over a $\pm$100 MHz window to derive accurate
sensitivities and baselines. 
The derived line parameters for the $l$-SiC$_3$ and $l$-SiC$_5$ species are given in Table \ref{line_parameters}.
The lines of $l$-SiC$_3$ and $l$-SiC$_5$ are shown in Fig. \ref{fig_sic3} and
\ref{fig_sic5}, respectively. 
The same method has been used for all the lines of 
the rhomboidal isomer of SiC$_3$ observed with the Yebes 40m radio telescope at 7\,mm,
and the IRAM 30m telescope at 3 and 2\,mm. The derived line parameters for this species are given in Table \ref{line_parameters_r} and the observed lines are
shown in Fig. \ref{fig_r-sic3}.

\begin{table}
\small
\caption{Derived line parameters for the lines of $l$-SiC$_3$ and $l$-SiC$_5$ in the Q band.} \label{line_parameters}
\centering
\begin{tabular}{ccccccccc}
\hline
Transition$^a$         & Frequency$^b$      &E$_{upp}$$^b$& $S^d$& $\int$ $T_A^*$ dv$^e$& T$_{cen}^f$& T$_{horn}^g$&$\sigma^h$& Notes\\
                  & (MHz)              & (K)     &   &(mK\,kms$^{-1}$)& (mK)& (mK)& (mK)& \\
\hline
$l$-SiC$_3$\\
6$_5$-5$_4$        & 32003.850$\pm$0.003&  6.4&  4.95 &               &     &     &0.12&A\\
6$_6$-5$_5$        & 32972.345$\pm$0.001&  5.3&  5.83 & 13.56$\pm$1.35& 0.43& 0.56&0.16& \\
6$_7$-5$_6$        & 33531.757$\pm$0.004&  5.9&  6.89 & 17.86$\pm$1.78& 0.57& 0.74&0.10& \\
7$_6$-6$_5$        & 37741.763$\pm$0.004&  8.2&  5.95 & 15.02$\pm$1.50& 0.48& 0.63&0.15& \\
7$_7$-6$_6$        & 38467.643$\pm$0.001&  7.2&  6.86 & 17.22$\pm$1.72& 0.55& 0.71&0.16& \\
7$_8$-6$_7$        & 38909.532$\pm$0.004&  7.8&  7.91 & 23.14$\pm$2.31& 0.79& 0.87&0.14&B\\
8$_7$-7$_6$        & 43403.260$\pm$0.004& 10.3&  6.95 & 15.18$\pm$1.51& 0.48& 0.63&0.20& \\
8$_8$-7$_7$        & 43962.898$\pm$0.002&  9.3&  7.88 & 20.50$\pm$2.05& 0.65& 0.85&0.18& \\
8$_9$-7$_8$        & 44319.288$\pm$0.005&  9.9&  8.92 & 15.58$\pm$1.55& 0.50& 0.65&0.20& \\
9$_8$-8$_7$        & 49015.957$\pm$0.005& 12.6&  7.96 & 14.42$\pm$1.44& 0.46& 0.60&0.30& \\
9$_9$-8$_8$        & 49458.104$\pm$0.002& 11.7&  8.89 & 15.16$\pm$1.51& 0.48& 0.63&0.39& \\
9$_{10}$-8$_9$     & 49750.661$\pm$0.005& 12.3&  9.93 & 15.09$\pm$1.50& 0.48& 0.63&0.15& \\
\hline
$l$-SiC$_5$\\
17$_{17}$-16$_{16}$& 31310.218$\pm$0.004& 13.4& 16.94 &  8.67$\pm$0.87& 0.24& 0.42& 0.11& \\
17$_{18}$-16$_{17}$& 31834.929$\pm$0.005& 14.3& 17.95 & 10.34$\pm$1.03& 0.29& 0.50& 0.11& \\
18$_{17}$-17$_{16}$& 32589.599$\pm$0.006& 17.2& 16.99 &  8.32$\pm$0.83& 0.21& 0.35& 0.09& \\
18$_{18}$-17$_{17}$& 33151.956$\pm$0.005& 15.0& 17.94 &               &     &     & 0.10&A\\
18$_{19}$-17$_{18}$& 33642.330$\pm$0.006& 15.9& 18.95 &  7.97$\pm$0.80& 0.23& 0.38& 0.10& \\
19$_{18}$-18$_{17}$& 34468.700$\pm$0.007& 18.9& 17.99 &  8.71$\pm$0.87& 0.25& 0.42& 0.09& \\
19$_{19}$-18$_{18}$& 34993.686$\pm$0.006& 16.7& 18.95 &  9.37$\pm$0.94& 0.27& 0.45& 0.11& \\
19$_{20}$-18$_{19}$& 35452.518$\pm$0.007& 17.6& 19.96 &               &     &     & 0.10&A\\
20$_{19}$-19$_{18}$& 36344.744$\pm$0.008& 20.6& 18.99 & 13.14$\pm$1.31& 0.37& 0.64& 0.09& \\
20$_{20}$-19$_{19}$& 36835.410$\pm$0.007& 18.5& 19.95 & 11.42$\pm$1.14& 0.32& 0.55& 0.10& \\
20$_{21}$-19$_{20}$& 37265.252$\pm$0.008& 19.4& 20.96 & 13.66$\pm$1.37& 0.39& 0.66& 0.11& \\
21$_{20}$-20$_{19}$& 38217.987$\pm$0.010& 22.4& 19.99 & 11.78$\pm$1.18& 0.33& 0.57& 0.14& \\
21$_{21}$-20$_{20}$& 38677.125$\pm$0.008& 20.3& 20.95 & 11.59$\pm$1.16& 0.33& 0.56& 0.10& \\
21$_{22}$-20$_{21}$& 39080.309$\pm$0.009& 21.3& 21.96 & 12.98$\pm$1.30& 0.37& 0.63& 0.11& \\
22$_{21}$-21$_{20}$& 40088.669$\pm$0.011& 24.4& 20.99 & 10.42$\pm$1.04& 0.30& 0.50& 0.11& \\
22$_{22}$-21$_{21}$& 40518.833$\pm$0.010& 22.3& 21.95 & 11.14$\pm$1.11& 0.31& 0.54& 0.10& \\
22$_{23}$-21$_{22}$& 40897.483$\pm$0.011& 23.3& 22.96 & 12.50$\pm$1.25& 0.35& 0.60& 0.12& \\
23$_{22}$-22$_{21}$& 41957.011$\pm$0.013& 26.4& 21.99 & 12.16$\pm$1.22& 0.34& 0.59& 0.12& \\
23$_{23}$-22$_{22}$& 42360.532$\pm$0.011& 24.3& 22.96 & 10.61$\pm$1.06& 0.30& 0.51& 0.10& \\
23$_{24}$-22$_{23}$& 42716.584$\pm$0.012& 25.3& 23.96 & 10.51$\pm$1.05& 0.30& 0.51& 0.11& \\
24$_{23}$-23$_{22}$& 43823.220$\pm$0.014& 28.5& 22.99 &               &     &     & 0.14&A\\
24$_{24}$-23$_{23}$& 44202.223$\pm$0.013& 26.4& 23.96 &               &     &     & 0.14&A\\
24$_{25}$-23$_{24}$& 44537.439$\pm$0.014& 27.5& 24.97 &  8.71$\pm$0.87& 0.25& 0.42& 0.14& \\
25$_{24}$-24$_{23}$& 45687.484$\pm$0.016& 30.7& 23.99 & 11.83$\pm$1.18& 0.33& 0.57& 0.16& \\
25$_{25}$-24$_{24}$& 46043.904$\pm$0.015& 28.6& 24.96 & 14.99$\pm$1.50& 0.48& 0.62& 0.18& \\
25$_{26}$-24$_{25}$& 46359.890$\pm$0.016& 29.7& 25.97 &  9.72$\pm$0.97& 0.27& 0.47& 0.17& \\
26$_{25}$-25$_{24}$& 47549.976$\pm$0.018& 32.9& 24.99 &               &     &     & 0.24&A\\
26$_{26}$-25$_{25}$& 47885.576$\pm$0.017& 30.9& 25.96 &               &     &     & 0.20&A\\
\hline
\hline
\end{tabular}
\tablefoot{
\tablefoottext{a}{Quantum numbers are $N$ (rotational quantum
number) and $J$ (total angular momentum, $J$ = $N$, $N$+1, $N$-1). They are
written as $N_J$.}
\tablefoottext{b}{Calculated frequency from the laboratory data
of \citet{McCarthy2000}. The adopted source velocity is of -26.5 kms$^{-1}$ \citep{Cernicharo2000}.}\\
\tablefoottext{c}{Energy of the upper level in K.}\\
\tablefoottext{d}{Line strength.}\\
\tablefoottext{e}{Integrated line intensity in mK kms$^{-1}$. Adopted uncertainty is 10\%.}\\
\tablefoottext{f}{Antenna temperature at line center in mK.}\\
\tablefoottext{g}{Antenna temperature at line horn in mK.}\\
\tablefoottext{h}{Measured data noise (1$\sigma$) in mK.}\\
\tablefoottext{A}{Heavily blended. Line parameters can not be derived.}\\
\tablefoottext{B}{Probably blended with a unidentified line.}
}
\end{table}

\normalsize
\section{Rhomboidal SiC$_3$}\label{app_r-SiC3}
The rhomboidal isomer of SiC$_3$ was discovered in IRC+10216 by \citet{Apponi1999a} from the observation of seven lines in the
3\,mm domain. Transitions of this species at 2mm were also observed by \citet{Cernicharo2000}. Here, we report the
detection of 45 rotational transitions of this
isomer with upper energy levels of up to 114\,K. Line parameters were derived in a similar way to
those of  $l$-SiC$_3$ and $l$-SiC$_5$ and are given in
Table \ref{line_parameters_r}. The rotational analysis of these data
is shown in Fig. \ref{fig_trot_r-sic3} (see below).

Most observed lines have an excellent S/N and exhibit U-shaped line profiles
(see Fig. \ref{fig_r-sic3}). This implies
that the emerging lines are formed in the external layers of the envelope. From the observed intensities we can estimate from a rotation diagram 
(see Fig. \ref{fig_trot_r-sic3}) the
rotational temperature and column density of the molecule.
A global fit to all the data provides T$_{rot}$=24.1$\pm$1.2\,K and N=(6.2$\pm$1.2)$\times$10$^{12}$ cm$^{-2}$.
The fit is shown as a dashed red line in Fig. \ref{fig_trot_r-sic3}. However, from the observed intensities it 
appears that for the lines with upper level energies below 50 K ($K_a$=0, 2, and 4) the rotational temperature is considerably
lower, T$_{rot}$=15.8$\pm$0.9\,K. The column
density for this gas component is N=(6.5$\pm$0.6)$\times$10$^{12}$ cm$^{-2}$. The fit is shown in Fig. \ref{fig_trot_r-sic3} as a blue line. The rotational
temperature for this gas component is similar to that
of $l$-SiC$_5$, SiC$_4$ and SiC$_6$.
Finally, a warmer gas component is found for transitions with energies of their
upper levels above 50\,K ($K_a\ge$4). The derived rotational temperature for this component is 44$\pm$8\,K and its column density
is (2.1$\pm$0.5)$\times$10$^{12}$ cm$^{-2}$ (purple line in Fig. \ref{fig_trot_r-sic3}). These results are in reasonable agreement with those derived by \citet{Apponi1999a} from a more reduced
number of lines and with lower sensitivities.

The lack of collisional rates for the SiC$_3$ isomers avoids a detailed study of their excitation conditions. However, as indicated previously by \citet{Apponi1999a}, the symmetry of $r$-SiC$_3$ and the fact that only $\Delta$$K_a$=0 radiative transitions are allowed, introduce some peculiarities in the population mechanisms of its energy levels. Collisions could connect all levels independently of the value of $K_a$, while radiation could connect only
levels within a K-ladder. Hence, we expect to have low
rotational temperatures within the $K$-ladders, and higher rotational temperatures across K-ladders. This is following what is represented in Fig. \ref{fig_trot_r-sic3} and the intensities provided in Table \ref{line_parameters_r}. The colors for the data points 
in this figure corresponds to
the different values of $K_a$. For each color, i.e., for each $K_a$,
the slope of the data is practically the same. The data for
different values of $K_a$ are shifted in energy (compare for example $K_a$=4 and $K_a$=6), but keep the same slope (rotational 
temperature). The data  for
$K_a$=0 and 2 can be fitted simultaneously and provide a rotational
temperature of 15.7$\pm$0.6\,K. For the $K_a$=4 and $K_a$=6 ladders the derived rotational temperature is of
17.8$\pm$1.4\,K and 18.6$\pm$1.6\,K, respectively. 
Hence, the average intra-ladder
rotational temperature is $\sim$17\,K, while the cross-ladder is $\sim$44\,K. The intra-ladder temperature is very similar to that
obtained for the rotational temperatures of the linear chains $l$-SiC$_3$, SiC$_4$, $l$-SiC$_5$, and SiC$_6$.

\begin{figure*}[t] \begin{center}
\includegraphics[width=0.8\textwidth]{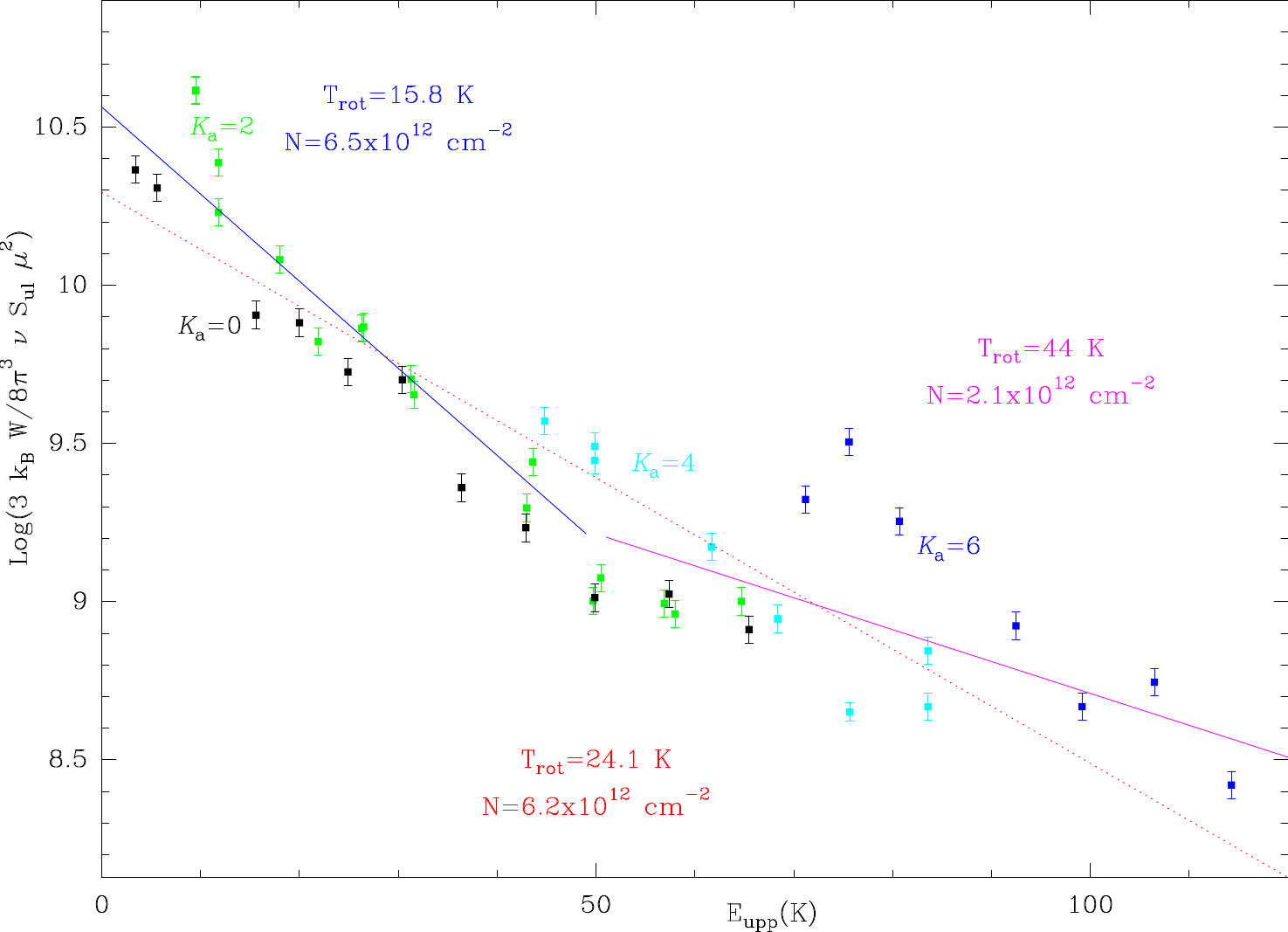}\end{center}  
  \caption{Rotational diagram for the lines of $r$-SiC$_3$. The blue line correspond to a fit
  to all transitions with energy of their upper level lower than 50\,K. The purple one corresponds to the
  fit to all data with upper level energies above 50\,K. The dashed red line corresponds to a global fit to
  all the data and represents an averaged temperature and column density. The data for each $K_a$ is represented by the same color (black
  for $K_a$=0, green for $K_a$=2, cyan for $K_a$=4 and blue for $K_a$=6.}         
   \label{fig_trot_r-sic3}
\end{figure*}

\begin{figure*}[t] \begin{center}
\includegraphics[width=1.0\textwidth]{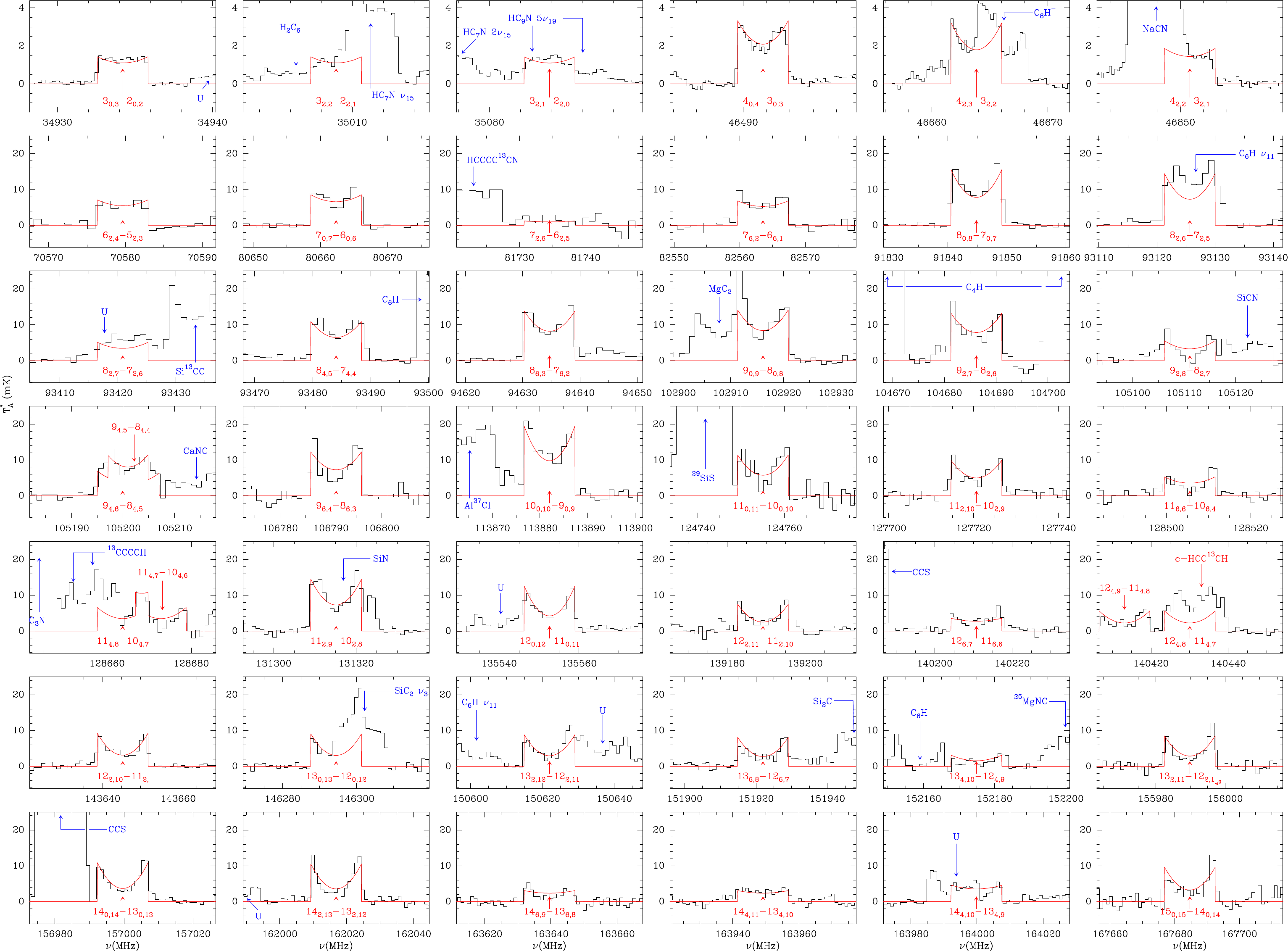}\end{center}  
  \caption{Lines of $r$-SiC$_3$ observed in this work at 7, 3, and 2mm. The red lines show the fitted line profiles (see text).
   Features from other species are labeled in blue. Abscissa and ordinate
    as in Fig. \ref{fig_sic3}.}         
   \label{fig_r-sic3}
\end{figure*}

\begin{table}
\small
\caption{Derived line parameters for the lines of rhomboidal SiC$_3$ at 7, 3, and 2 mm.} \label{line_parameters_r}
\centering
\begin{tabular}{ccccccccc}
\hline
Transition$^a$       & Frequency$^b$      &E$_{upp}$$^b$& $S^d$& $\int$ $T_A^*$ dv$^e$& T$_{cen}^f$& T$_{horn}^g$&$\sigma^h$& Notes\\
                   & (MHz)              & (K)     &   &(mK\,kms$^{-1}$)& (mK)& (mK)& (mK)& \\
\hline
\hline
$3_{0,3}-2_{0,2}$         &    34934.203$\pm$0.001&  3.4 &  3.00&  34.3$\pm$3.4   & 1.10  &   1.42&0.10 &\\  
$3_{2,2}-2_{2,1}$         &    35008.949$\pm$0.001&  9.5 &  1.67&  34.2$\pm$3.4   & 1.10  &   1.44&0.10 &A\\  
$3_{2,1}-2_{2,0}$         &    35083.884$\pm$0.001&  9.5 &  1.67&  34.2$\pm$3.4   & 1.10  &   1.49&0.10 &A\\  
$4_{0,4}-3_{0,3}$         &    46491.687$\pm$0.001&  5.6 &  4.00&  70.9$\pm$7.0   & 2.09  &   3.35&0.16 &\\  
$4_{2,3}-3_{2,2}$         &    46663.838$\pm$0.001& 11.8 &  3.00&  64.2$\pm$6.4   & 1.79  &   3.20&0.17 &\\  
$4_{2,2}-3_{2,1}$         &    46850.811$\pm$0.001& 11.8 &  3.00&  44.8$\pm$4.4   & 1.44  &   1.85&0.18 &\\  
$6_{2,4}-5_{2,3}$         &    70579.646$\pm$0.002&  18.0&  5.33& 169.9$\pm$16.9  & 5.47  &   7.19&2.00 &\\  
$7_{0,7}-6_{0,6}$         &    80662.322$\pm$0.003&  15.6&  6.98& 204.4$\pm$20.4  & 6.58  &   8.53&0.67 &\\
$7_{2,6}-6_{2,5}$         &    81540.082$\pm$0.002&  21.8&  6.43&                  &       &       &0.89 &B\\
$7_{6,2}-6_{6,1}$         &    81734.532$\pm$0.004&  71.2&  1.86&  29.3$\pm$5.5   & 1.77  &   2.30&0.96 &C\\  
$7_{6,1}-6_{6,0}$         &    81734.532$\pm$0.004&  71.2&  1.86&                  &       &       &0.96 &D\\ 
$7_{4,4}-6_{4,3}$         &    81777.540$\pm$0.002&  40.4&  4.72&                  &       &       &0.96 &B\\ 
$7_{4,3}-6_{4,2}$         &    81777.859$\pm$0.002&  40.4&  4.72&                  &       &       &0.96 &B\\ 
$7_{2,5}-6_{2,4}$         &    82563.549$\pm$0.003&  21.9&  6.43& 163.4$\pm$16.3  & 5.26  &   6.84&0.83 & \\ 
$8_{0,8}-7_{0,7}$         &    91844.820$\pm$0.004&  20.0&  7.98& 293.7$\pm$29.3  & 7.81  &  15.61&0.51 & \\
$8_{2,7}-7_{2,6}$         &    93125.636$\pm$0.002&  26.3&  7.50& 273.6$\pm$27.3  & 7.27  &  14.54&0.44 &A\\
$8_{6,3}-7_{6,2}$         &    93420.830$\pm$0.004&  75.6&  3.50& 112.3$\pm$11.2  & 3.41  &   5.11&0.42 & \\
$8_{6,2}-7_{6,1}$         &    93420.830$\pm$0.004&  75.6&  3.50&                  &       &       &0.42 &D\\
$8_{4,5}-7_{4,4}$         &    93484.045$\pm$0.003&  44.8&  6.00& 224.4$\pm$22.4  & 6.45  &  10.96&0.42 & \\
$8_{4,4}-7_{4,3}$         &    93484.920$\pm$0.003&  44.8&  6.00&                  &       &       &0.42 &D\\
$8_{2,6}-7_{2,5}$         &    94634.686$\pm$0.004&  26.5&  7.50& 285.1$\pm$28.5  & 8.19  &  13.92&0.56 & \\
$9_{0,9}-8_{0,8}$         &   102916.115$\pm$0.004&  24.9&  8.97& 291.2$\pm$29.1  & 8.37  &  14.22&0.71 & \\ 
$9_{2,8}-8_{2,7}$         &   104686.180$\pm$0.002&  31.3&  8.55& 272.0$\pm$27.2  & 7.82  &  13.28&1.00 & \\ 
$9_{6,4}-8_{6,3}$         &   105111.104$\pm$0.005&  80.7&  5.00& 114.1$\pm$11.4  & 3.28  &   5.57&1.00 & \\ 
$9_{6,3}-8_{6,2}$         &   105111.104$\pm$0.005&  80.7&  5.00&                  &       &       &1.00 &D\\
$9_{4,6}-8_{4,5}$         &   105199.949$\pm$0.003&  49.9&  7.22& 142.5$\pm$14.2  & 4.10  &   6.96&1.00 & \\
$9_{4,5}-8_{4,4}$         &   105202.048$\pm$0.003&  49.9&  7.22& 128.3$\pm$12.8  & 4.54  &  11.50&1.00 & \\
$9_{2,7}-8_{2,6}$         &   106791.028$\pm$0.004&  31.6&  8.55& 252.6$\pm$25.2  & 7.26  &  12.34&1.10 & \\
$10_{0,10}- 9_{0,9}$      &   113882.012$\pm$0.005&  30.4&  9.96& 368.5$\pm$36.8  & 9.81  &  19.58&1.50 & \\
$11_{0,11}-10_{0,10}$     &   124755.032$\pm$0.006&  36.4& 10.95& 216.2$\pm$21.6  & 5.75  &  11.49&2.00&\\
$11_{2,10}-10_{2,9}$      &   127720.669$\pm$0.003&  43.0& 10.63& 187.7$\pm$18.7  & 4.99  &   9.97&1.10&\\
$11_{6,6}-10_{6,5}$       &   128505.569$\pm$0.005&  92.5&  7.73& 116.6$\pm$11.6  & 3.54  &   5.31&1.10&\\
$11_{6,5}-10_{6,4}$       &   128505.570$\pm$0.005&  92.5&  7.73&                  &       &       &1.10&D\\
$11_{4,8}-10_{4,7}$       &   128663.730$\pm$0.003&  61.7&  9.55& 128.7$\pm$12.9  & 3.49  &   6.63&1.20&\\
$11_{4,7}-10_{4,6}$       &   128672.788$\pm$0.003&  61.7&  9.55& 128.7$\pm$12.9  & 3.49  &   6.63&1.20&\\
$11_{2,9}-10_{2,8}$       &   131315.143$\pm$0.005&  43.6& 10.64& 273.9$\pm$27.3  & 7.28  &  14.55&0.70&\\
$12_{0,12}-11_{0,11}$     &   135552.619$\pm$0.007&  42.9& 11.95& 199.0$\pm$19.9  & 4.23  &  12.69&0.51&\\
$12_{2,11}-11_{2,10}$     &   139189.193$\pm$0.003&  49.7& 11.66& 118.4$\pm$11.8  & 2.50  &   7.55&0.78&\\
$12_{6,7}-11_{6,6}$       &   140210.760$\pm$0.006&  99.2&  9.00&  85.5$\pm$8.5   & 2.75  &   3.58&0.81&\\
$12_{6,6}-11_{6,5}$       &   140210.763$\pm$0.005&  99.2&  9.00&                  &       &       &0.81&D\\
$12_{4,9}-11_{4,8}$       &   140413.124$\pm$0.003&  68.4& 10.67&  96.1$\pm$9.6   & 2.27  &   5.67&0.73&\\ 
$12_{4,8}-11_{4,7}$       &   140430.055$\pm$0.003&  68.4& 10.67&  96.1$\pm$9.6   & 2.27  &   5.67&0.73&\\ 
$12_{2,10}-11_{2,9}$      &   143645.312$\pm$0.006&  50.5& 11.67& 146.1$\pm$14.6  & 3.10  &   9.31&0.58&\\
$13_{0,13}-12_{0,12}$     &   146294.448$\pm$0.007&  49.9& 12.94& 143.4$\pm$14.3  & 3.05  &   9.14&0.67&\\
$13_{2,12}-12_{2,11}$     &   150621.962$\pm$0.004&  56.9& 12.68& 139.3$\pm$13.9  & 2.96  &   8.88&0.70&\\
$13_{6,8}-12_{6,7}$       &   151921.925$\pm$0.006& 106.5& 10.23& 128.2$\pm$12.8  & 2.72  &   8.17&0.82&\\
$13_{6,7}-12_{6,6}$       &   151921.933$\pm$0.006& 106.5& 10.23&                  &       &       &0.82&D\\ 
$13_{4,10}-12_{4,9}$      &   152174.858$\pm$0.003&  75.7& 11.77&  59.4$\pm$3.9   & 1.57  &   3.14&0.85&\\
$13_{4,9}-12_{4,8}$       &   152204.839$\pm$0.003&  75.7& 11.77&                  &       &       &0.85&B\\ 
$13_{2,11}-12_{2,10}$     &   155989.821$\pm$0.006&  58.0& 12.70& 134.5$\pm$13.4  & 2.86  &   8.57&0.91&\\
$14_{0,14}-13_{0,13}$     &   156999.640$\pm$0.007&  57.4& 13.94& 171.8$\pm$17.1  & 3.65  &  10.95&0.60&\\
$14_{2,13}-13_{2,12}$     &   162016.850$\pm$0.003&  64.7& 13.70& 165.6$\pm$16.6  & 3.54  &  10.50&0.94&\\
$14_{6,9}-13_{6,8}$       &   163639.571$\pm$0.006& 114.3& 11.43&  73.3$\pm$7.3   & 2.36  &   3.00&0.78&\\
$14_{6,8}-13_{6,7}$       &   163639.589$\pm$0.006& 114.3& 11.43&                  &       &       &0.78&D\\ 
$14_{4,11}-13_{4,10}$     &   163949.103$\pm$0.004&  83.6& 12.86&  73.1$\pm$ 7.3  & 2.35  &   3.16&0.95&\\
$14_{4,10}-13_{4,9}$      &   163999.813$\pm$0.004&  83.6& 12.86& 109.7$\pm$10.9  & 3.53  &   4.59&0.95&\\
$15_{0,15}-14_{0,14}$     &   167684.621$\pm$0.007&  65.5& 14.93& 151.9$\pm$15.1  & 3.23  &   9.68&1.20&\\
\hline
\end{tabular}
\tablefoot{
\tablefoottext{a}{Quantum numbers are $J, K_a$, and $K_c$.}
\tablefoottext{b}{Calculated frequency from the laboratory data
of \citet{McCarthy2000}. The adopted source velocity is of -26.5 kms$^{-1}$ \citep{Cernicharo2000}.}\\
\tablefoottext{c}{Energy of the upper level in K.} 
\tablefoottext{d}{Line strength.}
\tablefoottext{e}{Integrated line intensity in mK kms$^{-1}$. Adopted uncertainty is 10\%.}
\tablefoottext{f}{Antenna temperature at line center in mK.}
\tablefoottext{g}{Antenna temperature at line horn in mK.}
\tablefoottext{h}{Measured data noise (1$\sigma$) in mK.}\\
\tablefoottext{A}{Line partially blended with another feature.}\\
\tablefoottext{B}{Heavily blended. Line parameters can not be derived.}\\
\tablefoottext{C}{Upper limit.}\\
\tablefoottext{D}{Unresolved doublet.}
}
\end{table}

\end{appendix}

\end{document}